# Twist-enabled Transmissive Metasurface with Co-polarized Geometric Phase

Jiusi Yu[1,2], Haitao Li[1], Shijie Kang[1,3], Dongyi Wang[4], Pengfei Zhao[2], Jiayu Fan[1], Boyang Qu[1,3], Jensen Li[2], Xiaoxiao Wu[1,3,*]

[1]*Modern Matter Laboratory and Advanced Materials Thrust, The Hong Kong University of Science and Technology (Guangzhou), Nansha, Guangzhou 511400, Guangdong, China*
[2]*Department of Physics, The Hong Kong University of Science and Technology, Clear Water Bay, Kowloon, Hong Kong, China*
[3]*Low Altitude Systems and Economy Research Institute, The Hong Kong University of Science and Technology (Guangzhou), Nansha, Guangzhou 511400, Guangdong, China*
[4]*Department of Physics, University of Hong Kong, Hong Kong, China*
[*]*To whom correspondence should be addressed. E-mail: xiaoxiaowu@hkust-gz.edu.cn*

## Abstract

Metasurfaces have offered unprecedented control over electromagnetic (EM) waves across a wide range of frequency spectrum by manipulating their phase, amplitude, and polarization at subwavelength scales. Full wavefront control using metasurfaces requires $2\pi$ phase modulation, which is essential for advanced optical and photonic engineering. Common approaches, such as the Pancharatnam-Berry (PB) phases and resonant phases, face stringent limitations: PB phases essentially depend on circular polarization conversion, while resonant phases are inherently narrowband and require a complex design process. To overcome these challenges, we propose a broadband metasurface with a co-polarized transmissive geometric phase that achieves $2\pi$ phase coverage while conserving the circular polarization of incident EM waves. This co-polarized phase is enabled by a local twist angle between the upper and lower metallic patterns, forming a branch cut in the parameter space determined by the twist angle and frequency. The branch cut connects phase singularities of opposite chirality, ensuring broadband $2\pi$ phase coverage. We experimentally validate the presence of the branch cut and demonstrate broadband generation of arbitrary orbital angular momentum (OAM) for co-polarized output. Our approach provides a versatile method for designing broadband metasurfaces without altering circular polarizations, paving the way for development of compact optical and photonic devices.

# 1 Introduction

The manipulation of phase and polarization of light, or more broadly, electromagnetic (EM) waves, is crucial in modern optics and photonics, [1-9] with extensive applications in imaging, [10, 11] sensing, [12] telecommunications, [13] and so on. Beyond bulky optical elements, metasurfaces, comprised of ultrathin layers engineered with well-designed microstructures, present substantially more compact and efficient solutions to manipulate EM waves. In fact, previous studies have highlighted the comprehensive capabilities of metasurfaces to manipulate EM waves from GHz microwaves to visible light, [14, 15] with functions including arbitrary polarization control, [16-18] perfect absorption, [19] holographic imaging, [20-22] anomalous beam steering, [23-25] and the generation of optical vortices. [26, 27]

To achieve such advanced wave-manipulating functions, a full $2\pi$ phase modulation is generally required in metasurface design. [28] However, the Pancharatnam-Berry (PB) phase, a geometric phase approach, often employed for metasurfaces to systematically achieve the $2\pi$ phase modulation, is essentially limited to spin-converted functions. [3, 22] This challenge arises because, in most metasurfaces, the PB phase is only imparted to the cross-polarized components of the output. [29] Moreover, although the PB phase is broadband, metasurfaces designed using this approach are not necessarily broadband. The actual working bandwidth in use is limited by the proportion of cross-polarized transmission through the metasurface which can acquire the PB phase. [30] On the other hand, the resonant phase approach, which utilizes the resonances of metasurface microstructures, allows efficient phase modulation without spin conversion or spin dependence. [31] However, it is inherently narrowband and faces challenges such as complex design and fabrication processes, as well as the phase distortion caused by absorption losses, particularly at higher frequencies. [8, 32, 33]

In this work, we propose and experimentally demonstrate a spin-conserved broadband metasurface with a co-polarized geometric phase. [34] The key strategy is to introduce a local twist angle between the upper and lower layers of the designed metasurface. By varying the twist angle for a full period, the metasurface achieves a complete $2\pi$ phase coverage in co-polarized transmission across a broadband frequency range. The strategy takes advantage of the phase branch cut emerging in parameter space formed by the frequency and twist angle. A Lorentzian model is employed to analyze the emergence of the branch cut. We further experimentally demonstrate an application of our metasurface, the broadband and spin-conserved generation of orbital angular momentums (OAMs) for EM waves. In general, our findings can help us manipulate and design structures with spin-conserved $2\pi$ phase coverage over a broadband frequency range, which can significantly expand the use of metasurface in controlling the phase of the co-polarized transmitted component.

## 2 Results

2.1 Design of Spin-conserved Metasurface

The metasurface is composed of three layers of metallic patterns, as shown in Figure 1(a), with dielectric spacers sandwiched between the metallic layers. On the upper and lower metallic layers, there are slot patterns of three-fold rotational ($C_3$) symmetry arranged in a triangular lattice, whose resonant modes are coupled through the circular holes on the middle metallic layer arranged in a honeycomb lattice. The slot patterns are comprised of a circle slot of radius $R_c$, three long slots of radius $R_1$, and three short slots of radius $R_2$. Here, $R_1 = R_{\text{ave}} + \delta R$, and $R_2 = R_{\text{ave}} - \delta R$, where $R_{\text{ave}}$ is the average radius and $\delta R$ is the deviation. We then introduce a local twist angle $\theta$ for the slot patterns on the upper metallic pattern, and unit cells of the initial state ($\theta = 0$) and twist state ($\theta \neq 0$) are shown in Figure 1(b). It should be noted that the twist only locally rotates the slot patterns on the upper metallic layer, that is, each slot pattern is rotated around its own center. There is no rotation on the lower and middle metallic layers for the metasurface. A schematic depiction of circular-polarized electromagnetic (EM) waves impinging on the metasurface is shown in Figure 1(c). With selected geometric parameters for demonstration purposes, we have calculated the circular-polarized transmissions in the frequency range 16.0−20.0 GHz and averaged them under varying twist angle $\theta$ from 0 to 120°, and the transmission spectra are plotted as the inset of Figure 1(c). Obviously, for both left- and right-circular polarized (LCP and RCP) incidences, the co-polarized transmissions dominate, while the cross-polarized transmissions are near zero for any twist angle $\theta$. This indicates that our structure effectively suppresses cross-polarized transmission. However, the co-polarized transmission phases through the metasurface exhibit significantly different behaviors for different frequencies when varying the twist angle $\theta$ for a full period (from 0 to 120°). For instance, in the frequency range of 16.0–20.0 GHz, the transmission phases Arg($t_{\text{LL}}$) for LCP incidence span the entire $2\pi$ range at certain frequencies, such as 17.0, 17.5, and 18.0 GHz. At these frequencies, this metasurface can achieve full wavefront control for co-polarized transmissions by simply varying the twist angle $\theta$. On the contrary, this full $2\pi$ coverage is not observed at frequencies like 16.5 and 18.5 GHz.

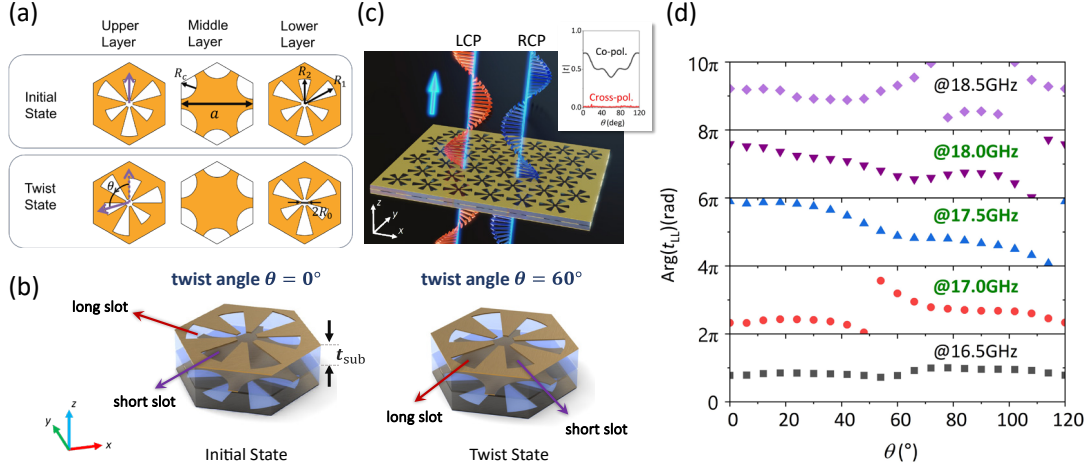

**Figure 1. Schematics of the metasurface and its transmission characteristics.** (a) Top-view schematic diagrams of the three metallic layers (local twist angle $\theta$), respectively. (b) Three-dimensional view of the unit cells with different twist angles, $\theta = 0°$ (initial state) and $\theta = 60°$ (twist state), respectively. (c) Overview of the proposed spin-conserved metasurface for circular-polarized EM waves, in which the co-polarized transmission dominates. The inset gives the average transmissions (from 16.0 to 20.0 GHz) for the selected geometric parameters with varying twist angle $\theta$, indicating near-zero cross-polarized transmissions (Co-pol. and Cross-pol. represent co-polarized transmission and cross-polarized transmission). (d) Phase spectra of co-polarized transmissions for left circular-polarized (LCP) incidence of EM waves. At certain frequencies (17.0, 17.5, and 18.0 GHz), the co-polarized transmission phases cover a range of $2\pi$, necessary for full wavefront control. The selected geometric parameters are $a = 6.0$ mm, $R_0 = 0.5$ mm, $R_1 = 2.7$ mm, $R_2 = 2.9$ mm, $R_c = 1.5$ mm, and $t_{sub} = 1.5$ mm. with a substrate permittivity of $\varepsilon_r = 3$.

## 2.2 Branch Cut and $2\pi$ Phase Coverage

To understand the underlying mechanism of the $2\pi$ phase coverage of our metasurface, we plot the transmission phase Arg($t_{LL}$) as color maps of the frequency $f$ and twist angle $\theta$. To discuss the broadband working frequency range, we vary the radius deviation $\delta R$ while keeping the average radius $R_{ave}$ constant. When the radius deviation $\delta R$ is small such as when $\delta R = 0.05$ mm, there is no such $2\pi$ phase coverage when varying the twist angle $\theta$, as shown in Figure 2(a). Meanwhile, after $\delta R$ becomes large enough, for example, when $\delta R = 0.10$ mm and 0.20 mm, two phase singularities emerge due to zero transmission when $\theta = 60°$ (equivalently, 180°),[35] as indicated in Figures 2(b) and 2(c). A branch cut thus emerges and connects the two phase singularities of opposite chirality in the parameter space that is spanned by the frequency $f$ and twist angle $\theta$. As a direct result, between the frequencies of the phase singularities, when varying the twist angle $\theta$ for a full period (from 0 to 120°), the transmission phase covers the $2\pi$ range, necessary for full manipulation of wavefront, such as the generation of OAM beams. Furthermore, it can be observed that the larger the radius deviation, the larger the frequency difference between the two phase singularities. This relationship leads to an expanded working frequency range for our metasurface when increasing the radius deviation.

To uncover the origin of the branch cut that is key to achieving $2\pi$ phase coverage, we employ the following Lorentzian model to fit the co-polarized transmission of our metasurface around the phase singularities

$$t = t_0 + \sum_{n=1}^{4} \frac{c_n \gamma_n \omega_n}{\omega^2 - \omega_n^2 - 2i\gamma_n \omega}. \tag{1}$$

Here, $t_0$ is the background transmission, $c_n$ ($n$ = 1, 2, 3, 4) are fitting coefficients, $\omega_n$ ($n$ = 1, 2, 3, 4) are eigenfrequencies (in angular frequency), and $\gamma_n$ are the corresponding damping coefficients. [36] Following numerical analysis of eigenmodes for our metasurface around 16.0–20.0 GHz, we employ four Lorentzian oscillators in the model. Both $\omega_n$ and $\gamma_n$ are directly retrieved from numerical analysis of the eigenmodes.

For the twist angle $\theta$ = 50°, 60°, and 70°, the simulated co-polarized transmissions and corresponding fitted lines are shown in Figures 2(d)-2(f), which show excellent agreement. The fitted model enables us to extend the frequency into the complex plane, allowing us to plot the co-polarized transmission phase Arg($t_{LL}$) in the complex frequency plane, as shown in Figures 2(g)-2(i). As can be seen, four poles always appear in the upper part of the complex frequency plane (since we adhere to the $e^{+i\omega t}$ time-harmonic convention), [37] corresponding to the four eigenmodes used in the fitting. When $\theta$ = 60°, there are two zeros on the real axis of the frequency, consistent with the transmission spectrum shown in Figure 2(e). However, despite the transmission spectra seeming similar when $\theta$ = 50° and 70°, the zeroes in the phases exhibit a pronounced difference in the complex frequency plane. As shown in both Figures 2(g) and 2(i), they move in opposite directions with respect to the real axis. Specifically, when $\theta$ = 50°, the first zero moves toward the upper half-plane, while the second zero moves toward the lower half-plane. Conversely, when $\theta$ = 70°, the reverse occurs. This opposite movement of the zeros in the complex frequency plane, either when the twist angle is smaller or larger ($\theta$ = 50° or 70°), and the frequency is lower or higher (around 16.8 GHz and 18.3 GHz), ensures that the phase singularities in the parameter space are associated with opposite chirality (see Note S1 in SI for detailed explanations) Thus, the branch cut emerges in the parameter space, closely related with those in the complex frequency plane.

In fact, when the twist angle $\theta = 50°$ or $70°$, the real frequency axis crosses the branch cut once and only once in the complex frequency plane. This crossing enables a $2\pi$ phase coverage when we vary the frequency between 16.0−20.0 GHz, as will be confirmed in Figure 3. Meanwhile, in the parameter space, as shown in Figures 2(b) and 2(c), when varying the twist angle $\theta$ for a full period (0-120°), it can result in the co-polarized transmission phase crossing the branch cut exactly once (see SI Note S12 for transmission amplitude in the parameter space). This scenario occurs for both LCP and RCP incidences when their frequency falls within the frequencies of the two phase singularities (see SI Note S2 for transmission phase $\text{Arg}(t_{RR})$ mapped in parameter space). In the working bandwidth, the change of the transmission phase exhibits a linear tendency, with the approximate relationship $\Delta\text{Arg}(t_{LL}) \approx -3\theta$ for LCP incidence, and $\Delta\text{Arg}(t_{RR}) \approx +3\theta$ for RCP incidence (see Note S3 in SI). We also emphasize that, despite the linear relationship is only an approximation, the change of the transmission phase is guaranteed to cover $2\pi$ when $\theta$ undergoes a periodic evolution (0 to 120°) owing to the branch cut in the parameter space.

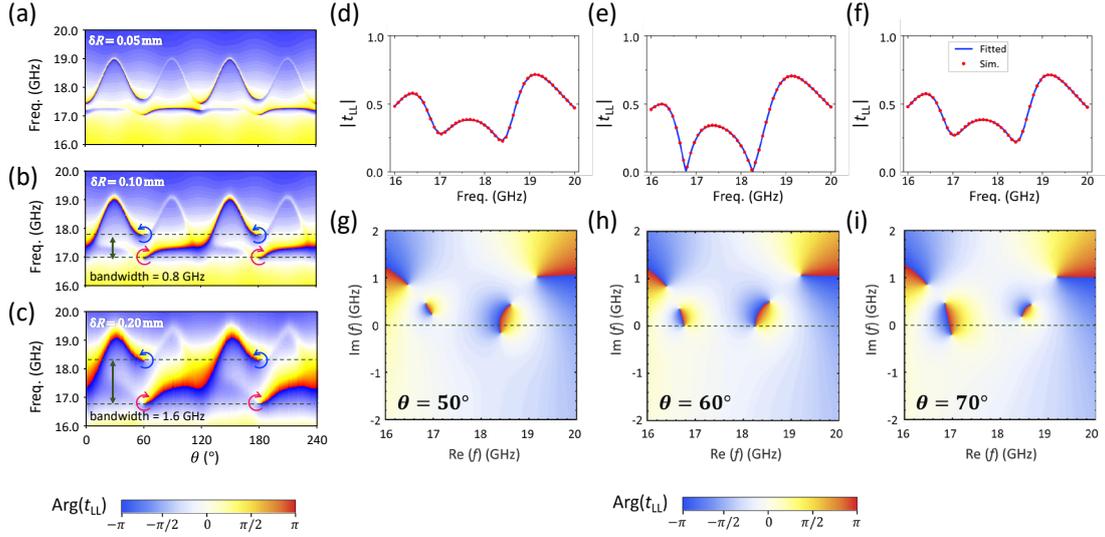

**Figure 2. Branch cut in the parameter space and phase singularity.** (a)-(c) Transmission phase $\text{Arg}(t_{LL})$ mapped in parameter space (frequency-twist angle $\theta$) for varying $\delta R$ values. The red and blue arrows encircle the phase singularity where transmission is zero, indicating two distinct directions of phase winding (clockwise direction and counterclockwise direction, respectively). The opposite phase singularities also determine the working bandwidth where the transmission phase covers $2\pi$ when varying twist angle $\theta$. (d)-(f) Simulated and fitted transmission spectra, which all agree excellently, when $\theta = 50°$(d), $\theta = 60°$ (e), and $\theta = 70°$ (f), respectively. (g)-(i) Fitted transmission spectra in complex frequency plane and corresponding singularities (poles and zeros), when $\theta = 50°$ (g), $\theta = 60°$ (h), and $\theta = 70°$

(i). Green dashed lines indicate the real frequency axis (Im(*f*) = 0). Branch cuts connected between the zeros and poles cross the real frequency axis when *f* = 16.7 GHz (*θ* = 50°) and *f* = 18.2 GHz (*θ* = 70°), respectively. In (d)-(i), *δR* = 0.20 mm.

2.3 Experimental Verification

A schematic diagram (not to scale) of the experimental setup for far-field measurements is shown in Figure 3(a). The fabricated metasurface samples, *θ* = 50° and 70° as shown in Figure 3(b), are measured to confirm the phase shifts of co-polarized transmissions. A metallic screen is used to support the sample and to reduce the impact of waves scattered by the edges (the specific sample structure and experimental setup can be found in Note S4 in SI). Both the emitter and receiver connected to the vector network analyzer (VNA) were linearly polarized horn antennas. After measuring the linearly polarized results, they are converted into circularly polarized transmissions based on the relationship

$$\begin{bmatrix} t_{LL} & t_{LR} \\ t_{RL} & t_{RR} \end{bmatrix} = \frac{1}{2} \begin{bmatrix} t_{xx} + t_{yy} + i(t_{xy} - t_{yx}) & t_{xx} - t_{yy} - i(t_{xy} + t_{yx}) \\ t_{xx} - t_{yy} + i(t_{xy} + t_{yx}) & t_{xx} + t_{yy} - i(t_{xy} - t_{yx}) \end{bmatrix},\quad(2)$$

where the first (second) sub-index represents the polarization state of the transmitted (incident) wave, and L and R correspond to LCP and RCP waves. [38]

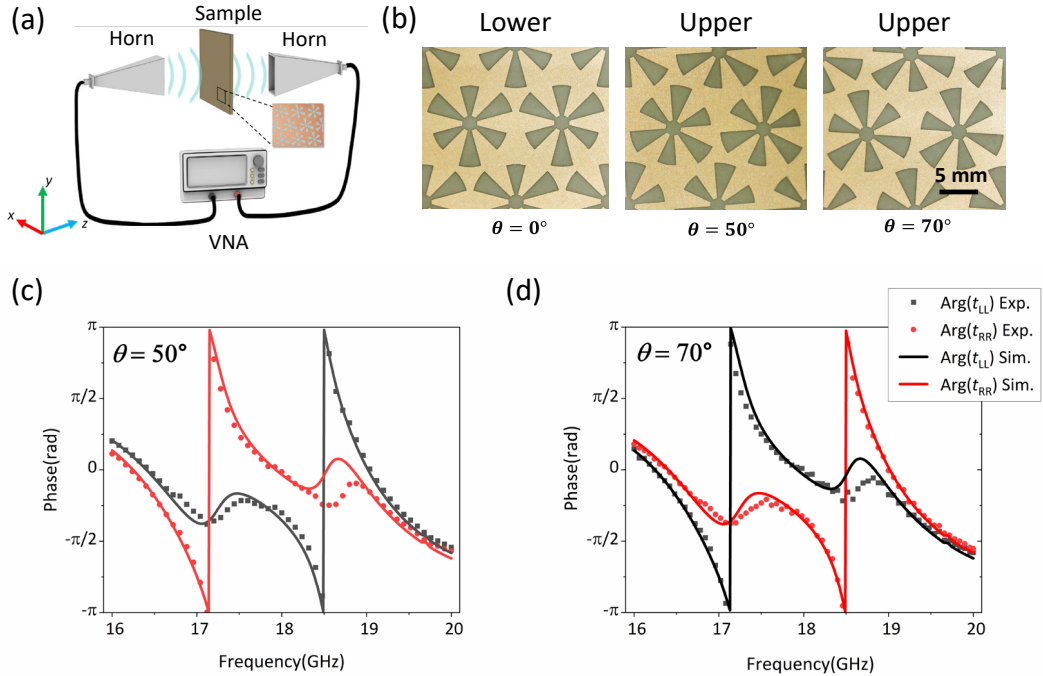

**Figure 3. Experimental comparisons and validation.** (a) Schematic diagram (not to scale) of the experimental setup of far-field measurements. Two horn antennas connected to the VNA are used as the emitter and receiver. (b) Photographs of the fabricated samples, including upper layers with different twist angles ($\theta = 50°$ and $\theta = 70°$ and lower layer ($\theta = 0°$ in all cases). (c), (d) Comparisons between simulations and experiments for different twist angles, including (c) $\theta = 50°$ and (d) $\theta = 70°$, respectively.

The measured co-polarized transmission phases, converted to a circular basis, are plotted in Figures 3(c) and 3(d), aligning excellently with simulations. Therefore, our metasurface exhibits spin-dependent transmission phases, that is, Arg($t_{LL}$) generally differs from Arg($t_{RR}$) when $\theta \neq 0$ or $60°$. Meanwhile, due to the mirror operation $M_x$ ($x \rightarrow -x$) which converts the twist angle $\theta$ to $120°-\theta$, we also have $t_{LL}(\theta) = t_{RR}(120°-\theta)$ (see SI Note S5). For example, as demonstrated in Figures 3(c) and 3(d), Arg($t_{LL}$) ($\theta = 50°$) differs from Arg($t_{RR}$) ($\theta = 50°$), while being equivalent to Arg($t_{RR}$) ($\theta = 70°$). At the same time, we have experimentally demonstrated that the transmission through our structure predominantly conserves the circular polarization, with virtually no cross-polarization transmission (for experimental results see Note S6 and Note S14 in SI). Moreover, phase wrappings (the abrupt jumps from $-\pi$ to $\pi$) are observed in both simulations and experiments, highlighting the essential role of branch cuts in achieving the desired phase manipulation. For example, the phase wrapping for LCP transmission occurs at either the lower or higher frequency (around 17.0 GHz and 18.5 GHz) around the transmission zeros, when the twist angle is either smaller or larger ($\theta = 50°$ or $70°$). This distinction experimentally showcases the branch cut in the parameter space, as illustrated in Figure 2(c). Meanwhile, we can see that for the selected twist angle, as we have a branch cut crossing the real axis of frequency, we indeed have a $2\pi$ phase coverage when the frequency crosses the transmission zeros,[28] corresponding to Figures 2(g) and 2(i). We further emphasize that for our metasurface, the cross-polarized transmissions are effectively suppressed, so the $2\pi$ phase coverage when we vary the local twist angle does not arise from the typical PB phases. In fact, the co-polarized transmission phase is acquired during the periodic evolution of the local twist angle $\theta$ suggesting that our metasurface exhibits a geometric-phase effect. Therefore, our scheme represents a promising approach for co-polarized wavefront engineering and the construction of compact optical and photonic devices (see Note S11 in SI for more discussion).

2.4 Generation of Orbital Angular Momentum
Since we have achieved the full $2\pi$ phase modulation, we now use the strategy to generate OAM beams of different orders in a broadband working frequency range to demonstrate the effectiveness of our scheme. The proposed broadband and spin-conserved OAM generator features a varying twisted angle $\theta$ at different positions, following the relation $\theta = l \times \alpha/3$, where $\alpha$ is the in-plane polar angle, and $l$ is topological charge of the arbitrary OAM we want to generate. Since we previously established an approximate linear relationship between the transmission phase and the twist angle, we can reasonably approximate the twist angle as linearly varying with the azimuth angle and use this assumption to design metasurfaces. As mentioned earlier, the metasurface we have discussed has a thickness of 1.5 mm. To optimize the transmission, we change the thickness $t_{sub}$ to 1.0 mm, such that the co-polarized transmission will increase significantly to around 0.9 in the working frequency range (see simulation results Note S7 and experimental results Note S14 in SI). The transmission of the optimized metasurface is then comparable to previous Huygens metasurfaces based on the PB phases or resonant phases.[36, 39-41] Due to the changes in the parameters of our structure, the distribution of the branch cut in the parameter space has also been altered (see Note S7 in SI). Here, we also take the LCP incidence for example. As depicted in Figure 4(a), with the varying twist angle $\theta = l \times \alpha/3$ for each unit cell, the transmitted LCP wave will carry an OAM of $+l$. On the other hand, for RCP incidence, the transmitted wave will carry an OAM of $-l$. Figure 4(b) shows the transmitted LCP wave for the LCP incidence, showcasing OAM manifested as the phase vortices and spiral patterns. As can be seen, the metasurface imparts OAM to the transmitted EM wave without altering its SAM, and the output field patterns carry corresponding OAM, $l = 1$ to 4, respectively (for the corresponding field intensity and phase distributions, see Note S13 in the SI). We also point out that, unlike resonant phases for co-polarized transmissions, our scheme based on the geometric phase offers strong robustness. In fact, the generation of OAM is essentially not affected by the local random perturbations on our metasurface (see Note S9 in SI).

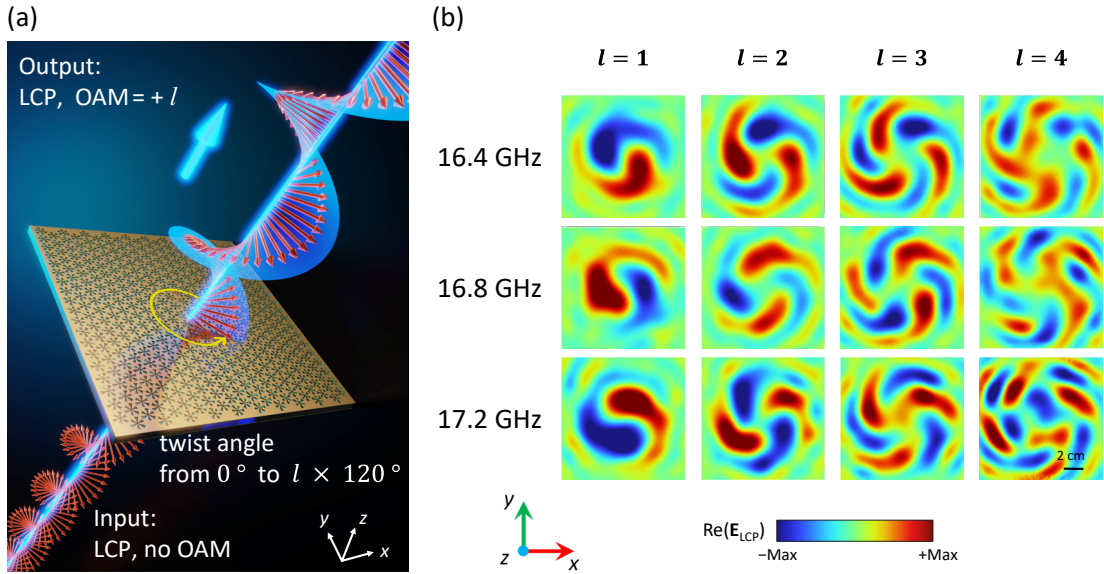

**Figure 4**. **Spin-conserved and broadband generation of OAM beams**. (a) Design of the OAM generator with our metasurface. The pattern on the upper metallic layer has a twist angle varying from 0 to $l \times 120°$, where $l$ is the designed OAM order. More specifically, the twist angle $\theta$ is proportional to the polar angle $\alpha$ (from 0 to 360° counterclockwise as indicated by the yellow arrow), that is, $\theta = l \times \alpha/3$. For LCP (RCP) incidence on the metasurface, the OAM beams carried by co-polarized output will be $+l$ ($-l$). (b) Simulated field maps for LCP incidence on the metasurface with designed OAM $l$ varying from 1 to 4 at certain frequencies (16.4, 16.8, and 17.2 GHz) in the working frequency range. To optimize transmission, we change $t_{sub}$ to 1.0 mm, while others are kept unchanged.

2.5 Experimental validation of orbital angular momentum generation

To experimentally validate the spin-conserved and broadband OAM generator based on our Huygens metasurface, we have also fabricated the samples based on the standard PCB technology. In the experiments as schematically depicted in Figure 5(a), a polarization-resolved probe is utilized for point-by-point imaging of the spatial electric field (specific photos of the experimental setup in Note S10 in SI). Consistent with previous experiments, a horn antenna is used to emit linearly polarized EM waves. By capturing the spatial electric field in four distinct polarization combinations, a comprehensive dataset of the electric field distribution can be collected. After performing these four sets of measurements, the linearly polarized components are converted into the circularly polarized. The co-polarized output for the OAM generator, which demonstrates vortex beam with OAM = +1 and OAM = +3 for LCP incidence, are shown in Figures 5(d) and 5(e), respectively. Meanwhile, the co-polarized output for RCP incidence demonstrates vortex beam with the opposite OAM (see Note S8 in SI). The output vortex beams and their associated OAM are also experimentally confirmed through the time-harmonic evolutions (see Movie S1 in SI). These experimental field maps, with output phase vortices aligning closely with the simulations, confirm that our samples indeed generate the co-polarized vortex beams with desired OAM values.

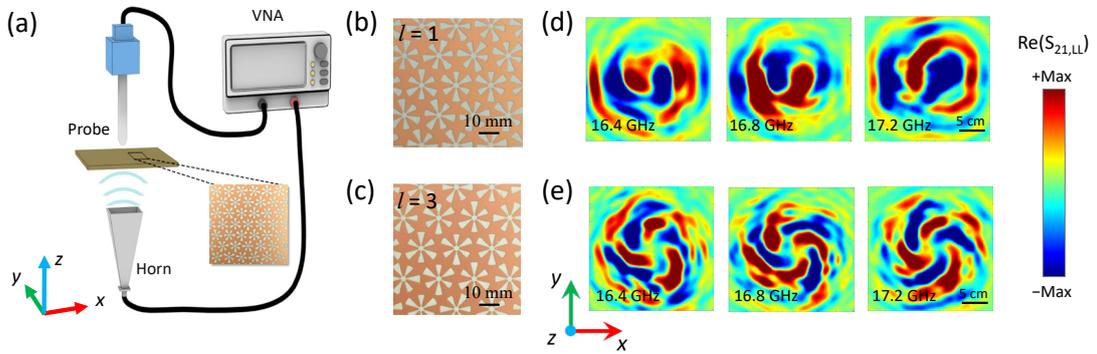

**Figure 5. Experimental imaging of generated co-polarized OAM beams.** (a) Experiment setup for imaging OAM beams with horn antenna as the source and probe

connected to VNA as the detector. The diagram is not to scale. (b), (c) Photographs of the fabricated OAM generator samples, with the designed OAM order $l = 1$ (b) and $l = 3$ (c). (d), (e) Measured $\mathbf{E}_{LCP}$ (extracted from experimental $S_{21}$ parameters obtained from VNA) distributions on an *x-y* plane 10 cm above the metasurface under measurement, shown at corresponding frequencies, for OAM = +1 (d) and OAM = +3 (e), respectively.

## 3 Discussion

In summary, we propose and experimentally demonstrate a scheme to achieve a spin-conserved Huygens metasurface with co-polarized geometric phase based on the branch cut in a parameter space. This scheme, through the introduction of a local twist angle to the metasurface, facilitates complete $2\pi$ phase coverage in the co-polarized transmission across a broad frequency range, without altering the circular polarization, or SAM, of the incident EM wave. The cross-polarized output is also effectively suppressed for the metasurface. Moreover, we have theoretically and experimentally demonstrated the existence of the branch cut in the parameter space and verified its essential role in the co-polarized geometric phase. By using this geometric phase, we designed and experimentally realized a broadband and spin-conserved OAM generator. Moreover, our scheme highlights the critical importance of parameter space in the design and analysis of metasurfaces. It provides a detailed understanding of the intricate relationship between the geometric phase and phase singularities. Since our scheme does not rely on specific material and geometric parameters, the co-polarized geometric phase is also applicable to higher frequency regimes, such as the terahertz and infrared. Therefore, our discovery allows for the effective and systematic design of spin-conserved metasurfaces, vital for advancing the development of more compact and sophisticated optical and photonic devices.

## Supplemental Material

See SI for supporting content.

## Acknowledgement

This research was supported by the National Natural Science Foundation of China (No. 12304348), Research Grants Council of Hong Kong (STG3/E-704/23-N,AoEP-502/20), Guangdong Provincial Project (2023QN10X059), Guangdong University Featured Innovation Program Project (2024KTSCX036), Guangzhou Municipal Science and Technology Project (No. 2024A04J4351), Guangzhou Higher Education Teaching Quality and Teaching Reform Engineering Project (2024YBJG087), HKUST - HKUST(GZ) 20 for 20 Cross-campus Collaborative Research Scheme (G011), and Research on HKUST(GZ) Practices (HKUST(GZ)-ROP2023021). J. Yu would like to thank Dr. Ruo-Yang Zhang for his comments and discussion.

## Author Competing Interest

The authors declare no conflict of interest.